\documentclass [12pt]{article}
\def\maj#1{\ifmmode\mbox{\usefont{U}{msb}{m}{n}#1}\else{\usefont{U}{msb}{m}{n}#1}\fi}
\def\v#1{\mathbf{#1}}
\usepackage[dvips]{graphics,epsfig,graphicx}
\usepackage{color}

\begin{document}

\title{\textbf{Many-body effects between \\ unbosonized
excitons}}
\author{M. Combescot and O. Betbeder-Matibet\\
\small{\textit{GPS, Universit\'e
Pierre et Marie Curie, CNRS,}}\\
\small{\textit{Campus Boucicaut, 140 rue de
Lourmel, 75015 Paris, France}}}
\date{}
\maketitle

\begin{abstract}
We here give a brief survey of our new many-body theory for
composite excitons, as well as some of the results we have
already obtained using it. In view of them, we conclude that,
in order to fully trust the results one finds, interacting
excitons should not be bosonized: Indeed, all effective bosonic
Hamiltonians (even the  hermitian ones !) can miss terms as
large as the ones they generate; they can even miss the dominant
term, as in problems dealing with optical
nonlinearities.
\end{abstract}

\newpage

We have recently developed a new many-body theory for
interacting composite bosons [1-8], which is actually an
important break-through: Indeed, up to now, the existing
many-body theories were appropriate to interactions between
fermions or bosons [9-11]. Although essentially all bosons known
in physics are composite bosons, there were no many-body
procedures adapted to them, most probably because it is far
more subtil to handle  composite bosons properly than to
handle true bosons or true fermions.

In order to ``do something'' in problems in which
interactions between composite bosons obviously play a role, it
is a common practice to replace composite bosons by true
bosons, provided that their interactions are ``dressed by
exchange'' -- which is considered as a way to take care of
their composite nature. The production of such effective
bosonic Hamiltonians is the aim of the wide literature on
bosonization [12,13].

This pragmatic idea is after all perfectly acceptable. The real
(major) trouble in semiconductor physics, is that the effective
bosonic Hamiltonian which has been produced [14,15] -- and
widely used -- to treat interactions between excitons, should
have been rejected long ago because it is not even hermitian:
This failure, which comes from unconsistent manipulations in
producing this effective Hamiltonian, has apparently been
missed by everyone for 35 years ! As we were the first to
point it out, it is easy to understand why our new theory for
interacting composite excitons has not been welcomed by the
experts in the field.

With a first line -- the Hamiltonian -- physically
unacceptable, there is no reason for the last line -- the
deduced results -- to be correct, except if some cancellations
in the missed terms miraculously take place. Unfortunately,
they don't ! Even if hermiticity is forced -- which is always
possible -- essentially the same amount of terms is missing. In
all the physical quantities we have studied up to now
using our new theory for composite excitons, the missed terms
are as large as the ones which appear. In some cases, in
particular the ones in which photons are involved,
\emph{i}.\
\emph{e}., in optical nonlinearities, the missed terms can
even be the dominant ones, as obvious from a dimensional
argument explained below. Excitons are deeply made of two
fermions; there is no way to get rid of their composite nature,
as we do when we bosonize them.

\vspace{0.5cm}

\noindent\textbf{Survey of our new many-body theory}

Our new many-body theory for composite bosons relies
on four nicely simple commutators [1,3]:
\begin{equation}
\left[H,B_i^\dag\right]=E_iB_i^\dag+V_i^\dag\ ,
\end{equation}
\begin{equation}
\left[V_i^\dag,B_j^\dag\right]=\sum_{mn}\xi_\mathrm{dir}
\left(_m^n\ _i^j\right)\,B_m^\dag B_n^\dag\ ,
\end{equation}
\begin{equation}
\left[B_m,B_i^\dag\right]=\delta_{m,i}-D_{mi}\ ,
\end{equation}
\begin{equation}
\left[D_{mi},B_j^\dag\right]=\sum_n\left[\lambda_h\left(_m^n\ 
_i^j\right)+\lambda_e\left(_m^n\ _i^j\right)\right]\, B_n^\dag\
.
\end{equation}
In the case of excitons, $H$ is the semiconductor Hamiltonian,
$B_i^\dag$ the creation operator of \emph{one} exciton in the
$i$ state, \emph{i}.\ \emph{e}., $(H-E_i)B_i^\dag|v\rangle=0$.
In terms of free electrons and holes, it reads
\begin{equation}
B_i^\dag=\sum_{\v k_e,\v k_h}\langle\v k_e,\v k_h|\phi_i\rangle
\,a_{\v k_e}^\dag b_{\v k_h}^\dag\ ,
\end{equation}
where $\langle\v k_e,\v k_h|\phi_i\rangle$ is the $i$ exciton
wave function in momentum space.

The above commutators generate two ``scatterings'':

\noindent (i) The first one $\xi_\mathrm{dir}\left(_m^n\
_i^j\right)$, which has the dimension of an energy, corresponds
to a direct Coulomb scattering between the ``in'' excitons
$(i,j)$ and the ``out'' excitons $(m,n)$, the excitons $m$ and
$i$ being made with the same electron-hole pair (see
fig.(1a)). It precisely reads
\begin{eqnarray}
\xi_\mathrm{dir}\left(_m^n\ _i^j\right)=\int d\v r_e\,d\v r_h\,
d\v r_{e'}\,d\v r_{h'}\,\phi_n^\ast(\v r_{e'},\v r_{h'})\,
\phi_m^\ast(\v r_e,\v r_h)\hspace{2cm}\nonumber\\
\left(V_{ee'}+V_{hh'}-V_{eh'}-V_{e'h}\right)\,\phi_i(\v r_e,\v
r_h)\,\phi_j(\v r_{e'},\v r_{h'})\ ,
\end{eqnarray}
where $\phi_i(\v r_e,\v r_h)$
is the $i$ exciton wavefunction in $\v r$ space and
$V_{ij}=e^2/|\v r_i-\v r_j|$.

\noindent (ii) The second scattering,
$\left[\lambda_h\left(_m^n\  _i^j\right)+\lambda_e\left(_m^n\
_i^j\right)\right]=2\lambda_{mnij}$, which is dimensionless, is
the sum of the possible carrier exchanges between the ``in''
excitons
$(i,j)$ and the ``out'' excitons
$(m,n)$, without any Coulomb process. In  $\lambda_h\left(_m^n\ 
_i^j\right)$, the excitons exchange their hole,
while in $\lambda_e\left(_m^n\ _i^j\right)$, they 
exchange their electrons (see fig.1b). $\lambda_h\left(_m^n\ 
_i^j\right)$ precisely reads
\begin{equation}
\lambda_h\left(_m^n\ _i^j\right)=\int d\v r_e\,d\v r_h\,
d\v r_{e'}\,d\v r_{h'}\,\phi_n^\ast(\v r_{e'},\v r_{h})\,
\phi_m^\ast(\v r_e,\v r_{h'})\,\phi_i(\v r_e,\v r_h)\,
\phi_j(\v r_{e'},\v r_{h'})\ ,
\end{equation}
while $\lambda_e\left(_m^n\ _i^j\right)=
\lambda_h\left(_n^m\ _i^j\right)=\lambda_h\left(_m^n\
_j^i\right)$.

With these four commutators, we can calculate any physical
quantities we wish. Indeed, physical quantities involving $N$
excitons can always be written in terms of matrix elements like
\begin{equation}
\langle v|B_{m_N}\ldots B_{m_1}\,f(H)\,B_{i_1}^\dag\ldots
B_{i_N} ^\dag|v\rangle\ .
\end{equation}
To calculate these matrix elements, we first push $f(H)$ to the
right, to end with $f(H)|v\rangle=f(0)|v\rangle$, if the vacuum
energy is taken as zero. This is done using eq.\ (1). In the
simplest case, $f(H)=H$, we have
\begin{equation}
H\,B_i^\dag=B_i^\dag\,(H+E_i)+V_i^\dag\ ,
\end{equation}
$V_i^\dag$ being then pushed to the right according to eq.\
(2). Another $f(H)$ of interest can be $(a-H)^{-1}$. It appears
in correlation effects or in optical nonlinearities, $a$ then
being $(\omega+i\eta)$. In this case, we use [4]
\begin{equation}
\frac{1}{a-H}\,B_i^\dag=B_i^\dag\,\frac{1}{a-(H+E_i)}+
\frac{1}{a-H}\,V_i^\dag\,\frac{1}{a-(H+E_i)}\ ,
\end{equation}
which follows from eq.\ (1), as easy to show by multiplying
eq.\ (10) by $(a-H)$ on the left and $(a-H-E_i)$ on the right.
A last $f(H)$ of interest is $e^{-iHt}$. It appears in problems
dealing with time evolution. We then use [16]
$$e^{-iHt}\,B_i^\dag=B_i^\dag\,e^{-i(H+E_i)t}+W_i^\dag(t)$$
\begin{equation}
W_i^\dag(t)=-\int_{-\infty}^{+\infty}\frac{dx}{2i\pi}\,
\frac{e^{-i(x+i\eta)t}}{x-H+i\eta}\,V_i^\dag\,\frac{1}
{x-H-E_i+i\eta}\ ,
\end{equation}
which follows from eq.\ (10) when inserted in the integral
representation of the exponential, namely
\begin{equation}
e^{-iHt}=-\int_{-\infty}^{+\infty}\frac{dx}{2i\pi}\,
\frac{e^{-i(x+i\eta)t}}{x-H+i\eta}\ ,
\end{equation}
valid for $t$ and $\eta$ positive.

Once $f(H)$ is pushed to the right, we start pushing the $B$'s
to the right, according to eqs.\ (3,4), to end with
$B_m|v\rangle=0$ and $D_{mi}|v\rangle=0$, which follows from
eq.\ (4) applied to $|v\rangle$.

In problems dealing with many excitons, most of them are
usually in the same state $0$. It is then convenient to note
that the iteration of eqs.\ (1,4) leads to [17]
\begin{equation}
\left[H,B_0^{\dag N}\right]=NE_0\,B_0^{\dag N}+NB_0^{\dag
N-1}V_0^\dag
+ \frac{N(N-1)}{2}\, B_0^{\dag
N-2}\sum_{mn}\xi_\mathrm{dir}\left(_m^n\ _0^0
\right)\,B_m^\dag B_n^\dag\ ,
\end{equation}
\begin{equation}
\left[B_m,B_0^{\dag N}\right]=N\,B_0^{\dag N-1}(\delta_{m,0}
-D_{m0})-N(N-1)\,B_0^{\dag N-2}\sum_p\lambda_h\left(_m^p\
_0^0\right)\,B_p^\dag
\ .
\end{equation}

The real -- challenging -- difficulties in problems dealing
with interacting excitons, is not so much the calculation of
these matrix elements, which after all, can be tedious but is
always conceptually trivial; the real difficulty is the
determination of the appropriate matrix elements we have to
calculate to get the physical quantity we want. Indeed, all our
background and practice in problems dealing with interactions,
rely on perturbation theory, \emph{i}.\ \emph{e}., on
the possibility to isolate an interacting
potential $V=H-H_0$, from the Hamiltonian. In the case of
composite excitons, this is not possible: We cannot isolate
from the Coulomb potential, the part corresponding to
interactions
\emph{between} two excitons. Indeed, if we see them as made
with $(e,h)$ and
$(e',h')$, we would say that the Coulomb potential between
these excitons is $(V_{ee'}+V_{hh'}-V_{eh'}-V_{e'h})$, while if
we see them as made with $(eh')$ and $(e'h)$, this potential
should be $(V_{ee'}+V_{hh'}-V_{eh}-V_{e'h'})$. Since the
electrons (holes) are undistinguishable, there is no way to
know ! Actually, the quantity $V_i^\dag$, defined in eq.\ (1),
plays the role of this interacting potential. It however has
the dimension of a  pair creation operator,
this is why we have called it ``Coulomb creation potential''. To
support this idea, we can note that the first terms of the
right hand side of eqs.\ (9,10,11), which contain $(H+E_i)$,
are in fact the ``zero order terms'' of the left hand side of
these equations, \emph{i}.\ \emph{e}., the terms in the absence
of Coulomb interaction between the exciton $i$ and the rest of
the system.

The real difficulty to get the physics of interacting excitons,
is thus to write the quantities we look for, not in terms
of
$V$, as we usually do, but in terms of the
Hamiltonian $H$. The Hamiltonian
\emph{a priori} contains all the necessary informations
on the interactions existing in the system. Consequently, even
if this may not appear as obvious at first, there should be a
way to write all quantities dealing with interacting excitons in
terms of $H$ only. (Note that, since there is no $V$, there is
no $H_0$ -- because its existence would mean that we
could define $V$ -- so that there is no
``unperturbed state''
$|0\rangle$).

As an example of such a possible new writing, we have recently
considered the lifetime and scattering rates of excitons [18].
In the usual case, when $H=H_0+V$, they are given by the Fermi
golden rule, which reads in terms of $V$. We have
shown that the lifetime of an initial state
$|\psi_0\rangle$ can also be written as
\begin {equation}
\frac {t}{\tau_0}=\langle\psi_0|\hat{H}|F_t(\hat{H})|^2
\hat{H}|\psi_0
\rangle\ ,
\end{equation}
where $\hat{H}=H-\langle\psi_0|H|\psi_0\rangle$, while $F_t(E)=
(e^{-iEt}-1)/E$, so that $|F_t(E)|^2=2\pi\,t\,\delta_t(E)$ with
$\delta_t(E)=(\pi\,E)^{-1}\sin(Et/2)$ being the usual delta
function of width $(2/t)$. It is easy to check that, for
$H=H_0+V$ and $|\psi_0\rangle =|0\rangle$, with
$(H-E_n)|n\rangle=0$, the Fermi golden rule follows from eq.\
(15): Indeed, as $\hat{H}|\psi_0\rangle =
(1-|0\rangle\langle0|)\ V|0\rangle$ is just $\sum_{n\neq 0}|n
\rangle\langle|V|0\rangle$, we do recover
\begin{equation}
\frac{1}{\tau_0}=2\pi\,\sum_{n\neq 0}\left|\langle n|V|0\rangle
\right|^2\,\delta_t(E_0-E_n)\ .
\end{equation}

\vspace{0.5cm}

\noindent\textbf{Diagrams associated to this new many-body
theory}

The construction of a new many-body theory has for final
goal, to write all processes in which enter
interactions, at any order. Of course, to reach this goal, we
always start in a pedestrian way, by generating the
corresponding terms, through some kind of recursion procedure,
in order to obtain their mathematical expressions. However, if
we go to high orders, most of the time, these expressions
become so complicated that we do not ``see'' their physical
meaning anymore. This is why many-body theories are usually
associated to diagrams which are just the visual representation
of these mathematical quantities. These diagrams are
really useful because they allow to ``see'' the physics involved
in the corresponding terms, in an easy way.

In our new many-body theory for interacting excitons, enter two
conceptually different scatterings. The direct Coulomb
scattering between two excitons $\xi_\mathrm{dir}\left(_m^n\ 
_i^j\right)$ is basically similar to the Coulomb scattering
between free carriers. It gives rise to diagrams which
look very much like the Feynman diagrams. On the opposite, the
Pauli scatterings, $\left[\lambda_h\left(_m^n\ 
_i^j\right)+\lambda_e\left(_m^n\ _i^j\right)\right]$, are the
conceptually new part of our theory. Although they appear as
scatterings between two excitons $(i,j)$, transforming
them into
$(m,n)$, they originate from Pauli exclusion, which
makes the excitons close-to-bosons only, as can be seen from
eqs.\ (3,4). However, Pauli exclusion being $N$-body
``at once'' --- an additional exciton having to be made with
carriers in states different from \emph{all} the ones involved
in the excitons already present --- it is clear that these Pauli
scatterings between two excitons only, cannot be the
quantities which appear at the end, in
physical effects in which
$N$ excitons are involved. Nevertheless, they are quite
convenient in the intermediate stages of the theory, because
they allow to write recursion relations between scalar products
of
$N,(N-1),(N-2),\cdots$\,exciton states.

Using eq.\ (14), we can in particular show that
\begin{eqnarray}
A_{mi}^{(N)}&=&\langle v|B_0^{N-1}B_mB_i^\dag B_0^{\dag N-1}
|v\rangle\nonumber
\\ &=&
\delta_{m,i}\,A_{00}^{(N-1)}+(N-1)C_{mi}^{(N)}
+(N-1)^2(N-2)^2\sum_{pj}\lambda_{00ip}\,
\lambda_{jm00}\,A_{pj}^{(N-2)},\hspace{0.5cm}
\end{eqnarray}
where $A_{00}^{(N)}=N!F_N$ has already been
calculated in references [5,7], while $C_{mi}^{(N)}$ is given by
\begin{equation}
C_{mi}^{(N)}=\delta_{0,i}A_{m0}^{(N-1)}-2\sum_p\lambda_{m0ip}
\,A_{p0}^{(N-1)}-(N-1)(N-2)\,\delta_{m,0}
\sum_p\lambda_{ooip}\,A_{p0}^{(N-2)}\ .
\end{equation}
If we iterate this equation and perform the sums over the
dumb indices $(p,j,\ldots)$, through closure relations, we end
with the ``skeleton Pauli diagrams'' [8] shown in fig.2.

This figure essentially says that the scalar product of one
exciton $m$ along with $(N-1)$ excitons $0$ on the left, and
one exciton $i$ along with $(N-1)$ excitons $0$ on the right,
can be separated into a term in which the excitons $0$ do not
play any role, and terms in which one, two, \ldots excitons $0$
out of $N$, exchange their carriers with the excitons $m$ and
$i$, in any possible way. The $N$-dependent prefactors are
just the number of ways to choose these excitons $0$ among
$(N-1)$, on each side, while the sign is related to the
parity of the number of exchanges involved in the diagram, as
usual.

This diagrammatic representation of the matrix element of
$N$-exciton states with one exciton different from $0$ on each
side, can be extended to any other scalar products [17] having
more than one exciton different from $0$. Although this
extension may appear as reasonable, it did not appear to
us obvious, at first ! It in fact relies on heavy
calculations which use recursion relations similar to the one
of eq.\ (17). This ``reasonable'' extension has actually been
made possible due to the introduction of the ``skeleton
Pauli diagrams'' [8], which allow to get rid of the relative
positions of the Pauli scatterings
$\lambda_{mnij}$, and which make disappearing all the  awkward
factors 2 due to
$\lambda_e\left(_m^n\ _0^0
\right)=\lambda_h\left(_m^n\ _0^0\right)=
\lambda_h\left(_n^m\ _0^0\right)$.

Using these ``skeleton Pauli diagrams'', in which can possibly
enter some additional direct Coulomb scatterings between two
exciton lines, due to possible $H$ contributions to the matrix
elements of interest, we can represent any
quantity we wish, dealing with interacting excitons,
in a transparent way.

\vspace{0.5cm}

\noindent\textbf{Some applications of this new many-body theory}

We have used this new many-body theory in a few problems
in which interacting excitons are involved:

\noindent 1) \textit{Link with the effective bosonic
Hamiltonian for excitons}

In order to make the link with the results obtained from the
usual effective bosonic Hamiltonian for excitons, we have shown
[1] that the scattering $V_{mnij}$ which appears in the
interaction term of this Hamiltonian,
$(1/2)\sum_{mnij}V_{mnij}\bar{B}
_m^\dag\bar{B}_n^\dag\bar{B}_i\bar{B}_j$, where the
$\bar{B}_i$'s are true boson operators,
$[\bar{B}_i,\bar{B}_j^\dag]=\delta_{i,j}$, is just
\begin{equation}
V_{mnij}=\xi_\mathrm{dir}\left(_m^n\ _i^j\right)-\sum_{pq}
\xi_\mathrm{dir}\left(_m^n\ _p^q\right)\lambda_{pqij}\ .
\end{equation}
Its replacement by
\begin{equation}
\xi_\mathrm{dir}\left(_m^n\
_i^j\right)-\frac{1}{2}\sum_{pq}\left[
\xi_\mathrm{dir}\left(_m^n\ _p^q\right)\lambda_{pqij}+\lambda_
{mnpq}\xi_\mathrm{dir}\left(_p^q\ _i^j\right)\right]\ ,
\end{equation}
would have allowed the effective Hamiltonian to be at least
hermitian.

This hermiticity is however not enough to get correct results
on interacting excitons, because $\xi_\mathrm{dir}$ and
$\lambda_{e,h}$ are conceptually independent scatterings: They
may appear in many other ways than the ones written in eqs.\
(19,20). In particular, in problems dealing with photons,
$\lambda_{e,h}$ can appear alone [16,19], so that the
corresponding terms are simply missed by \emph{any} effective
bosonic Hamiltonian, whatever these scatterings are. Indeed, 
$V_{mnij}$ must have the dimension of an energy, while
$\lambda_{e,h}$ is dimensionless. Moreover, Coulomb processes
which may enter these physical effects, have to appear with
energy denominators, in order to have the same dimension as the
pure Pauli terms, which contain $\lambda_{e,h}$ only. In
problems dealing with photons, these energies are in fact the
exciton detunings, so that the Coulomb processes are negligible
in front of the pure Pauli terms -- missed by the effective
bosonic Hamiltonians -- in the large detuning limit.

\noindent 2) \textit{Scattering rates and lifetime of excitons}

We have shown that the lifetime of an exciton state in which
all the excitons are in the same state $0$, is related to the
scattering rates towards other exciton states in which two
excitons differ from $0$, by the relation [18]
\begin{equation}
\frac{1}{\tau_0}=\alpha\,\sum_{(i,j)\mathrm{couples}}\frac{1}
{T_{ij}}\ ,
\end{equation}
where $\alpha=1$ if the excitons are bosonized and $\alpha=1/2$
if their composite nature is kept.
This shows that it is impossible to find effective scatterings
for the effective bosonic Hamiltonian, giving both the lifetime
and scattering rates correctly.

This type of sum rule comes from closure relation. We have
shown that, while the one for bosonized excitons is of course
the standard closure relation for elementary particles,
\begin{equation}
I=\frac{1}{N!}\sum_{i_1,\ldots,i_N}\bar{B}_{i_1}^\dag\cdots
\bar{B}_{i_N}^\dag|v\rangle\langle v|\bar{B}_{i_N}\cdots\bar
{B}_{i_1}\ ,
\end{equation}
a closure relation also exists for composite excitons, in spite
of the fact that these states form a non-orthogonal
overcomplete set [20]. It actually reads [21]
\begin{equation}
I=\frac{1}{(N!)^2}\sum_{i_1,\ldots,i_N}B_{i_1}^\dag\cdots
B_{i_N}^\dag|v\rangle\langle v|B_{i_N}\cdots
B_{i_1}\ .
\end{equation}
Consequently, all sum rules will appear differently for exact
and boson excitons, making the corresponding quantities
irretrievably different.

\noindent 3) \textit{The trion as an exciton interacting with a
carrier}

This new many-body theory for excitons interacting with
excitons can be adapted to excitons interacting with free
carriers [22-24].

The simplest of these problems is surely the trion: The exciton
interacts with one electron either by Coulomb interaction or by
electron exchange. Using this approach, we can calculate the
trion oscillator strength in a quite easy way [25]. We have
shown that it is one exciton volume divided by one sample
volume smaller than the exciton one, making the trion
impossible to see in the linear response of a macroscopic
semiconductor sample.

This approach should allow us to face the much harder problem
of the linear response of a doped semiconductor quantum well.
Our old work on this problem [26] is quite
unsatisfactory, not at all because Coulomb screening is
missing, but because the electron-electron repulsion has
been neglected. In order to include it, \emph{i}.\ \emph{e}.,
in order to study the rearrangement of a Fermi sea to the
sudden appearance of an exciton -- not simply to a hole, as we
did -- we have to handle not only Coulomb interaction
between the exciton and the Fermi sea electrons, but also all
possible electron exchanges. This again makes our new many-body
theory for interacting excitons rather unavoidable.

\noindent 4) \textit{Optical nonlinearities}

The optical nonlinearities come from interactions with the
virtual excitations induced in the matter by the photons. In
the case of semiconductors, these excitations are the excitons.
Consequently, in order to derive the optical nonlinearities in
semiconductors properly, it is necessary to know how to handle
interactions between excitons in a secure way. This is why our
new many-body theory for composite excitons is going to be all
the most useful for semiconductor optical nonlinearities.

We have first reconsidered the calculation of the third
order susceptibility $\chi^{(3)}$ [16]. This calculation
contains a major difficulty, unsolved before our work, which
comes from the fact that $\chi^{(3)}$ appears as the sum of 16
terms which all increase linearly with the sample volume. Of
course, as the susceptibility is an intensive quantity, these
volume-linear terms have to cancel. However, this cancellation
has to be shown very carefully in order to get the correct
volume-free terms which remain once the volume-linear terms are
removed. Our many-body theory allows to calculate
these 16 terms explicitly and to combine them in order to show
that the volume-linear terms do disappear as expected. However,
this clean cancellation in fact generates volume-free terms
missed by the previous approaches, which come from pure Pauli
interactions between the virtual excitons coupled to the
photons. Again, these missed pure Pauli terms are the dominant
ones at large detuning.

Another nonlinear effect in semiconductor physics is the
exciton optical Stark effect [27]. The seeds of our new
many-body theory were actually in our old works on this effect,
in particular the Coulomb creation potential $V_i^\dag$. The
introduction of the two scatterings $\xi_\mathrm{dir}$ and
$\lambda_{e,h}$ however make the study of this exciton optical
Stark effect quite easy, in particular if we are interested in
subtle polarization effects.

Our many-body theory is also very convenient to study spin
manipulations by laser pulse [19]. We have recently
predicted the spin precession of an electron bound to an
impurity, when the sample is irradiated by a pump beam tuned in
the transparency region. This precession, which is due to
exchanges between the electron bound to the impurity and
the electron of the virtual exciton coupled to the pump photon,
lasts as long as the laser pulse, so that it can be used for
ultrafast spin manipulation.

\vspace{0.5cm}

\noindent\textbf{Conclusion}

\begin{itemize}
\item We have developed a new many-body theory adapted to
interactions between composite excitons.

\item It relies on two ``scatterings'': One is associated to
Coulomb processes, the other is associated to carrier exchanges
induced by Pauli exclusion. This Pauli scattering can appear
alone or mixed with Coulomb processes, in many more subtle ways
than the one appearing in the effective bosonic Hamiltonian for
excitons. When it appears alone, the corresponding terms are
\emph{a priori} missed by the effective bosonic Hamiltonians,
whatever are the scatterings they use. These missed terms are
often the dominant ones, due to dimensional arguments.

\item We have generated a diagrammatic representation of this
theory. The direct Coulomb scattering gives rise to diagrams
similar to the Feynman diagrams. On the opposite,
the Pauli scattering gives rise to diagrams which are
conceptually new: Pauli exclusion being $N$-body ``at once'',
the corresponding ``skeleton Pauli diagrams'' show all
possible carrier exchanges between 2,3,4\ldots excitons, in a
quite transparent way.
\item This new many-body theory is going to be of great
interest for all problems dealing with interactions between
excitons and between excitons and free carriers, as well as for
all optical nonlinearities in semiconductors, because, for the
first time, it allows a clean treatment of the possible carrier
exchanges.
\item This new theory is also going to be of importance in many
other fields than semiconductor physics, because essentially
all particles considered as bosons, are in fact composite
bosons.

\end{itemize}

\vspace{0.5cm}

\hbox to \hsize {\hfill REFERENCES
\hfill}

\noindent
[1] M. Combescot and O. Betbeder-Matibet, Europhys.\ Lett.\
\textbf{58}, 87 (2002).

\noindent
[2] M. Combescot and O. Betbeder-Matibet, Europhys.\ Lett.\
\textbf{62}, 140 (2003).

\noindent
[3] O. Betbeder-Matibet and M. Combescot, Eur.\ Phys.\ J.\ B
\textbf{27}, 505 (2002).

\noindent
[4] M. Combescot and O. Betbeder-Matibet,  Europhys.\ Lett.\
\textbf{59}, 579 (2002).

\noindent
[5] M. Combescot and C. Tanguy, Europhys.\ Lett.\ \textbf{55},
390 (2001).

\noindent
[6] M. Combescot and C. Tanguy, Europhys.\ Lett.\ \textbf{63},
787 (2003).

\noindent
[7] M. Combescot, X. Leyronas and C. Tanguy, Eur.\ Phys.\ J.\ B
\textbf{31}, 17 (2003).

\noindent
[8] M. Combescot and O. Betbeder-Matibet, Cond-mat/0403073.

\noindent
[9] A. Abrikosov, L. Gorkov and I. Dzyaloshinski, Methods of
quantum field theory in statistical physics, Prentice-Hall,
Englewood Cliffs, N.J. (1963).

\noindent
[10] A. Fetter and J. Walecka, Quantum theory of many-particle
systems, Mc. Graw-Hill (1971).

\noindent
[11] G. Mahan, Many-particle Physics, Plenum Press, N.Y.
(1990).

\noindent
[12] T. Usui, Prog.\ Theor.\ Phys.\ \textbf{23}, 787 (1957).

\noindent
[13] For a review, see A. Klein and E.R. Marshalek, Rev.\
Mod.\ Phys.\ \textbf{63}, 375 (1991).

\noindent
[14] E. Hanamura and H. Haug, Phys.\ Report \textbf{33}, 209
(1977).

\noindent
[15] H. Haug and S. Schmitt-Rink, Prog.\ Quantum Electr.\
\textbf{9}, 3 (1984).

\noindent
[16] M. Combescot, O. Betbeder-Matibet, K. Cho and H. Ajiki,
Cond-mat/0311387.

\noindent
[17] O. Betbeder-Matibet and M. Combescot, Eur.\ Phys.\ J.\ B
\textbf{31}, 517 (2003)

\noindent
[18] M. Combescot and O. Betbeder-Matibet, Phys.\ Rev.\ Lett.\
\textbf{93}, 016403 (2004).

\noindent
[19] M. Combescot and O. Betbeder-Matibet, Cond-mat/0404744, to
appear in Solid State Com.\ (2004).

\noindent
[20] This had already been pointed out by M. Girardeau, Jour.\
Math.\ Phys.\ \textbf{4}, 1096 (1963).

\noindent
[21] M. Combescot and O. Betbeder-Matibet, Cond-mat/0406045.

\noindent
[22] M. Combescot, Eur.\ Phys.\ J.\ B \textbf{33}, 311 (2003).

\noindent
[23] M. Combescot and O. Betbeder-Matibet, Solid State Com.\
\textbf{126}, 687 (2003).

\noindent
[24] M. Combescot, O. Betbeder-Matibet and F. Dubin,
Cond-mat/0402441.

\noindent
[25] M. Combescot and J. Tribollet, Solid State Com.\
\textbf{128}, 273 (2003).

\noindent
[26] M. Combescot and P. Nozi\`{e}res, J.\ Phys.\ (Paris)
\textbf{32}, 913 (1972).

\noindent
[27] For a review, see M. Combescot, Phys.\ Report
\textbf{221}, 168 (1992).

\newpage

\begin{figure}
\centerline{ \scalebox{0.7}{\includegraphics{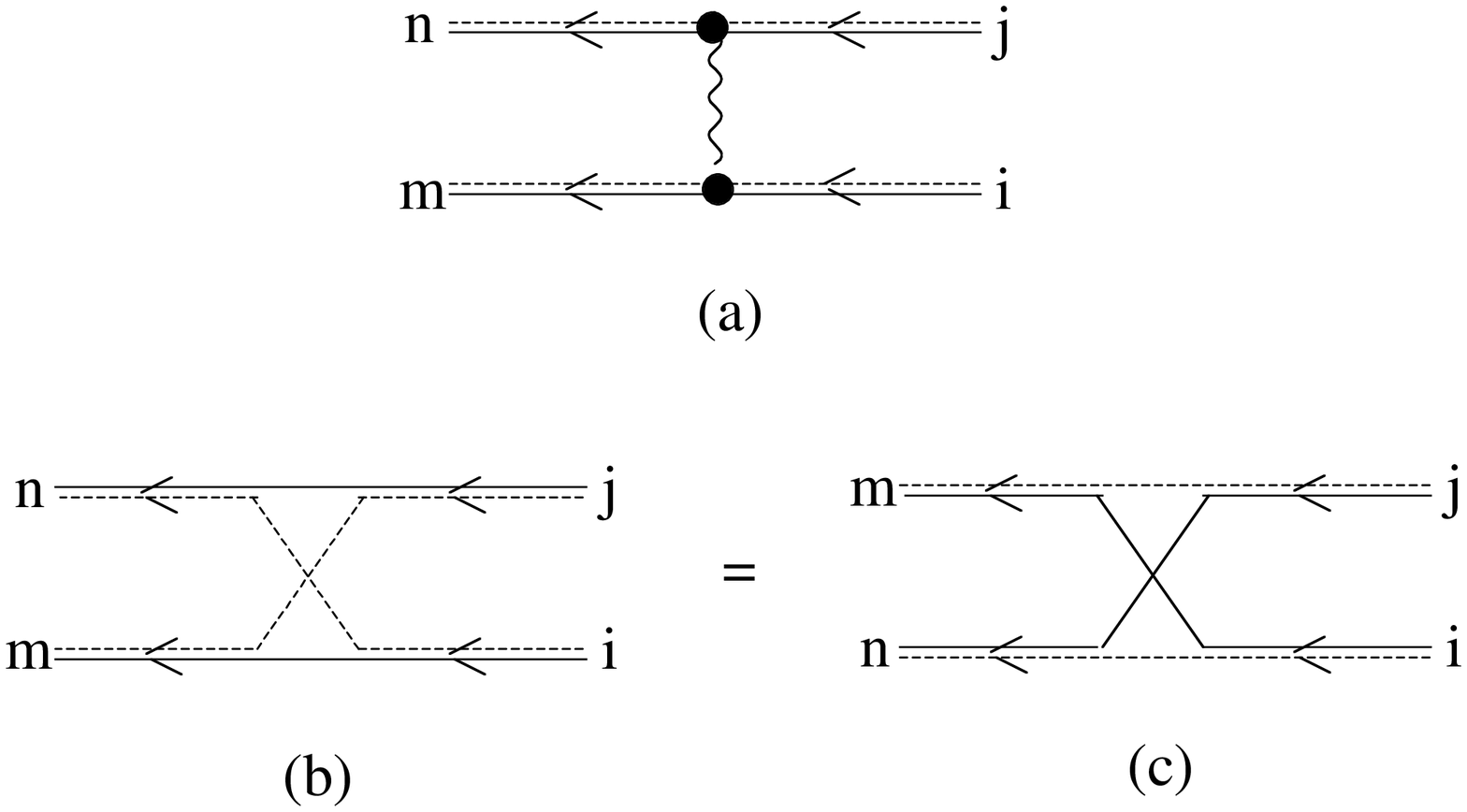}}}
\caption{\footnotesize{
(a) Direct Coulomb scattering $\xi_\mathrm{dir}\left(_m^n\
_i^j\right)$ between the ``in'' excitons $(i,j)$ and the
``out'' excitons $(m,n)$. (b) Exchange scattering $\lambda_h
\left(_m^n\ _i^j\right)$ in which the excitons exchange their
holes. In (c), which shows $\lambda_e\left(_n^m\ _i^j\right)$,
they exchange their electrons. Note that 
$\lambda_h\left(_m^n\ _i^j\right)=
\lambda_e\left(_n^m\ _i^j\right)$.}}
\end{figure}

\clearpage

\begin{figure}
\centerline{ \scalebox{0.7}{\includegraphics{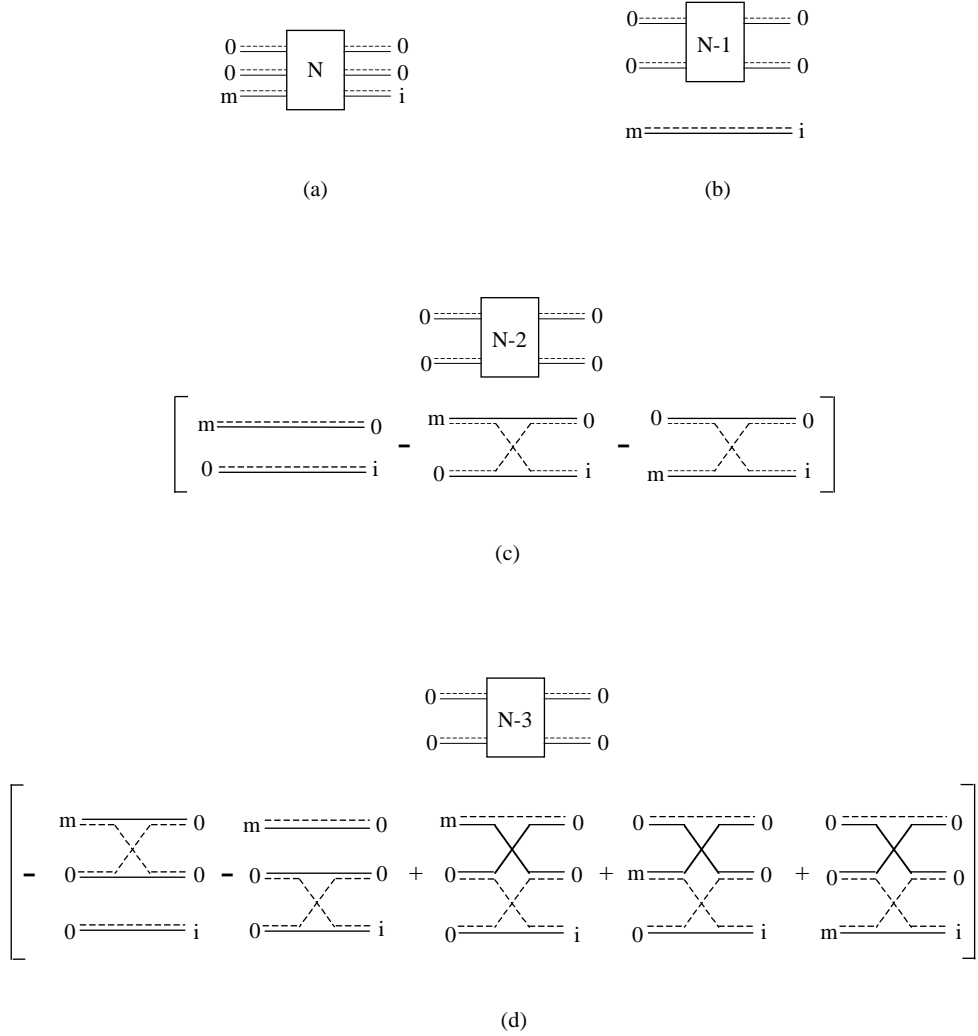}}}
\caption{\footnotesize{
Diagrammatic expansion in ``skeleton Pauli diagrams'' of the
scalar product $A_{mi}^{(N)}$, defined in
eq.\ (17) and shown in (a), in which all excitons but one
are in the same state
$0$. In (b), the
$(m,i)$ excitons do not exchange any carrier with the sea of
$(N-1)$ excitons $0$. In (c), one exciton 0 is involved in
carrier exchanges with $(m,i)$. In (d), two excitons $0$ are
involved; and so on\ldots The term in which
$p$ excitons $0$ are involved is equal to the scalar product,
$A_{00}^{(N-1-p)}$, of the remaining $(N-1-p)$ excitons $0$,
shown as a box with ($N-1-p$) inside, multiplied by all
possible carrier exchanges between the
$(m,i)$ excitons and the $p$ excitons $0$, topologically
connected or not, but without diagrams with excitons $0$ alone.
The prefactor, not shown in the figure, is the number of ways to
choose the $p$ excitons
$0$ among $(N-1)$, on each side, \emph{i}.\ \emph{e}.,
$[(N-1)!/(N-1-p)!]^2$, the sign being linked to the parity of
the number of carrier exchanges, as usual.}}
\end{figure}

\clearpage

\end{document}